# A Review Study of NIST Statistical Test Suite: Development of an indigenous Computer Package

J K M Sadique Uz Zaman and Ranjan Ghosh
Institute of Radio Physics and Electronics, University of Calcutta
92, Acharya Prafulla Chandra Road, Kolkata – 700 009
jkm_sadique@rediffmail.com,   rghosh47@yahoo.co.in

*Abstract*

*A review study of NIST Statistical Test Suite is undertaken with a motivation to understand all its test algorithms and to write their C codes independently without looking at various sites mentioned in the NIST document. All the codes are tested with the test data given in the NIST document and excellent agreements have been found. The codes have been put together in a package executable in MS Windows platform. Based on the package, exhaustive test runs are executed on three PRNGs, e.g. LCG by Park & Miller, LCG by Knuth and BBSG.  Our findings support the present belief that BBSG is a better PRNG than the other two.*

**1. Introduction:**

The Statistical Test Suite developed by NIST [1] is an excellent and exhaustive document looking at various aspects of randomness in a long sequence of bits. The Test Suite was developed after DES was cracked in order to choose today's AES. It is a very important tool to understand randomness not only of the PRNGs but also of the crypto ciphers. The document provides many links [2,3,4]. In some cases [2,3] there are different types of useful information regarding different modules used for their programs and in some other cases [4] there are large data set of random bit arrays obtained from different PRNGs. Instead of endeavoring to study the data and information available in those links, initiatives have taken to enrich understanding of all the test algorithms with a belief that a capability would duly take shape to develop indigenous codes with scopes of future improvements, if possible.  Had it been that an initiative would have been taken towards understanding the coding methodology of NIST test programs, the computing system issues would have taken the front seat for a long time keeping the scientific issues at the back.

The NIST has documented 15 statistical tests and in each test it adopted first a procedure to find the statistic of chi-square variation ($\chi^2$) of a particular parameter for the given bit sequence with that obtained from the theoretical studies of an identical sequence under the assumption of randomness. It then adopted a technique to transform the $\chi^2$ data to a randomness probability data, named as P-value.  Techniques adopted for conversion of the $\chi^2$ data to respective P-value are described in Sec. 2. In Sec.3, 15 tests algorithms are narrated with better clarity of understanding. Indigenous codes of all the test algorithms have been developed and have been taken together in a computer package. The package has gives scopes to the user to select a particular test or few tests together or all tests. The package has been used to study the randomness features of three PRNGs. Among them, two are Linear

Congruent Generators, one (LCG-PM) by Park & Miller [5,6] and another (LCG-K) by Knuth [7] and the third one is the Blum Blum Shab Generator (BBSG) [8]. The results and discussion in connection to the test runs for the three PRNGs are presented in Sec. 4. The conclusion in brief is in Sec. 5.

## 2. Conversion of Test Statistic data to Probability value(s):

NIST has adopted broadly two approaches to convert the chi-square variation ($\chi^2$) to P-value. In first approach the P-value is calculated based on the $\chi^2$ data as one parameter, while in second approach the P-value is calculated based on the degrees of freedom (K) data as the first parameter and the $\chi^2$ data as the second parameter. Six tests 1, 3, 6, 9, 13 and 15 do belong to the first approach, while the rest nine tests 2, 4, 5, 6, 7, 8, 10, 11, 12 and 14 belong to the second one. The name of each test is given in Table A.

For the first approach, a particular statistic parameter is observed for the entire bit sequence together and its $\chi^2$ variations are computed against the theoretical values obtained for the same parameter considering a corresponding bit sequence under the assumption of randomness – the $\chi^2$ data are converted to P-value considering Standard Normal (Gaussian) distribution function, $\Phi(x)$ where x is related to the $\chi^2$ data. In the second approach, the entire bit sequence is divided into N blocks and a concept of (K+1) classes with K degrees of freedom is introduced based on theoretical studies of a particular statistic parameter desired for the respective tests considering a corresponding identical bit sequence under the assumption of randomness. The said parameter is observed block-wise across the entire bit sequence and its $\chi^2$ data are computed against the block-wise theoretical values of the parameter – the $\chi^2$ data are then converted to P-value considering gamma function, $\Gamma(a, x)$, where the parameter a is related to the degrees of freedom (K) and the parameter x, to the $\chi^2$ data.

The Gaussian distribution function is given below as,

$$\Phi(z) = 1/\sqrt{(2\pi)} \left(\int e^{-x^2/2} dx\right), \text{ integrated from } x = -\infty \text{ to } x = z.$$

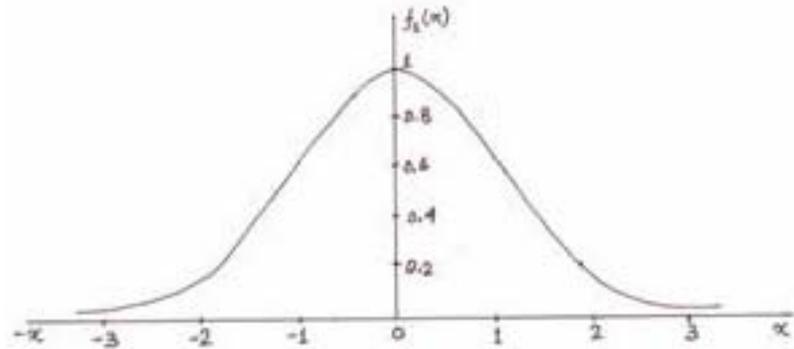

Fig. 1: Plot of $f_1(x) = \exp(-x^2/2)$

Fig. 1 shows the plot of $f_1(x)$ leading to Gaussian function. It may be noted that $\Phi(z)$ is unity when $f_1(x)$ is integrated from $x = -\infty$ to $x = +\infty$. Hence $\Phi(z)$ when $f_1(x)$ is integrated from

x = -∞ to x = +z is related to $\chi^2$ data since z = ½($\chi^2$); and P-value = 1 - Φ(z). The larger is the $\chi^2$ variation, the larger is the value of Φ(x) and lesser is the value of P-value indicating that the bit sequence in consideration can be termed as non-random if P-value becomes smaller than a threshold value.

It may be noted that Φ(z) is used only in test 13 where the parameter z = ½($\chi^2$) assumes positive as well as negative values; hence Φ(z) is evaluated within limits of +ve and –ve values and the P-value is obtained by subtracting the result of integration from unity.

If z in the Gaussian distribution assumes positive values, one resorts to error function given as,

$$\text{erf}(z) = (2/\sqrt{\Pi})(\int e^{-x^2} dx), \text{ integrated from } x = 0 \text{ to } x = z \text{ (less than } +\infty\text{)}.$$

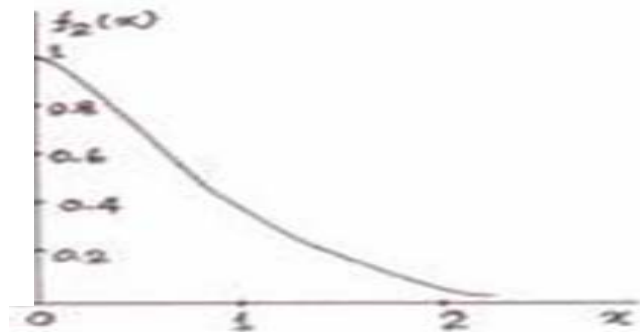

Fig. 2: Plot of $f_2(x) = \exp(-x^2)$

Fig. 2 shows a plot of $f_2(x)$ leading to error function. If the limit of integration varies from 0 to +∞, the result of integration is unity. The error function is used whenever the parameter z connected to $\chi^2$ variations assumes always positive values. It may be noted that erf(z) can be derived from Φ(z). It is readily observed that if the parameter x derived under the assumption of Gaussian distribution is to be used in error function, it is to be divided by √2. For tests 1, 3, 6, 9 and 15, z assumes positive value and P-value is calculated as [1 – erf(z)].

For the rest nine tests 2, 4, 5, 7, 8, 10, 11, 12, and 14, different approach of computation is taken to evaluate the P-value. A concept of degrees of freedom is introduced in these tests in the form of blocks or classes. For such cases, instead of adopting the Gaussian distribution function or error function, one resorts to a distribution function based on gamma function which has two parameters. The gamma function Γ (a, x) is given below as,

$$\Gamma(a, z) = \int x^{a-1} e^{-x} dx, \text{ where the limit of integration is from } x=0 \text{ to } x=z$$

Here the **a** is related to the degrees of freedom and **z** is related to statistic $\chi^2$ variation.

Fig. 3 below shows plots of Γ (a, z) for few values of **a**. The P-value is computed as,

$$\text{P-value} = 1 - \Gamma(a, z) / \Gamma(a, \infty)$$

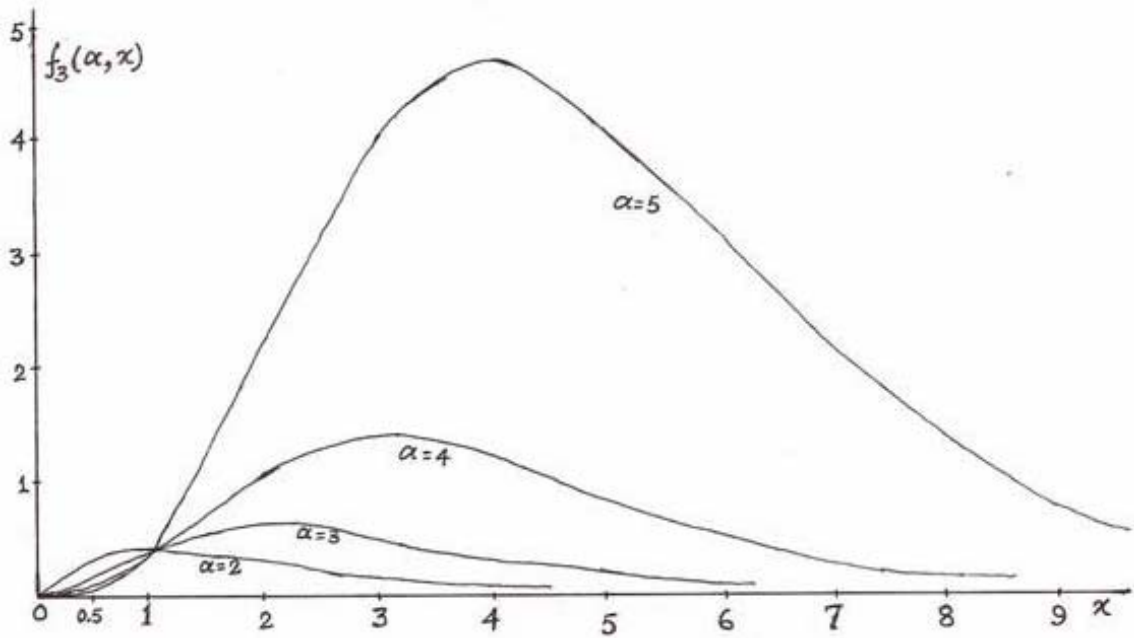

Fig. 3: Plot of $f_3(a, x) = x^{a-1}\exp(-x)$

where $\Gamma(a, z)$ is the integration of gamma function from $x = 0$ to $x = z$ for a particular value **a**. $\Gamma(a, \infty)$ is the integration of same function from $x = 0$ to $x = +\infty$ for the same value of **a**.

## 3. Description of Test Algorithms:

The available theoretical studies related to many statistic parameters of bit sequences under the assumption of randomness are the computational basis to estimate the $\chi^2$ variation. From broad theoretical considerations, the 15 tests can be categorized into four categories, namely Frequency Tests (tests: 1 – 4), Test for Repetitive Patterns (tests: 5 and 6), Tests for Pattern Matching (tests: 7 – 12) and Tests based on Random Walk (tests 13 – 15). The algorithms of tests 1, 3, 6, 9, 13 and 15 do consider the entire bit sequence together for computation of $\chi^2$ variation and computes the P-value based on error function, expect test 13 for which Standard Normal (Gaussian) function is used. The algorithms of tests 2 and 7 divide the entire bit sequence in N blocks and compute the P-value based on gamma function using N as the degrees of freedom. The algorithms of tests 4, 5, 8 and 10 divide the entire bit sequence in N blocks and also consider (K+1) classes obtained from respective theoretical studies and computes the P-value based on gamma function using K as the degrees of freedom, instead of N. The algorithms of tests 11, 12 and 14, without dividing the bit sequence into blocks, introduce (K+1) classes obtained from respective theoretical studies and computes the P-value based on gamma function with K degrees of freedom. The algorithms of all the 15 tests are described below in sub-sections 3.1 through 3.15.

### 3.1. Test 1: The Frequency (Mono-bit) Test (1 P-value with 1 parameter, x)

**Prelude**:

(1) Through this test it is intended to see if the frequencies of 1 and 0 across the entire n-bit sequence are approximately equal meaning that the proportion of each of 1s and 0s is close to ½.
(2) If the number of 0s and 1s are not the same, it is intended to see if their difference falls within the limit of randomness.

**Background information in respect of randomness:**

(1) This test is derived from central limit theorem for the random number.
(2) The classic De Moivre-Laplace theorem states that for a large number of trials the distribution of binomial sum, normalized by $\sqrt{n}$, is closely approximate by a standard normal distribution.

**Focus of Computation**:

(1) Each bit 0 and 1 in the string is represented by -1 and 1 respectively by using the mathematical relation $X_i = 2\acute{\varepsilon}_i - 1$, where $X_i$ represents new value of the bit $\acute{\varepsilon}_i$ at the $i^{th}$ position.
(2) The sum of $X_i$ represents $S_n$ and $S_{obs} = |S_n|/\sqrt{n}$.
(3) $x = S_{obs}/\sqrt{2}$.
(4) P-value = 1 - erf(x).

### 3.2. Test 2: Frequency Test within a Block (1 P-value with 2 parameters, a & x)

**Prelude**:

(1) One can note that even if the first half on the n-bit sequence is full of 1 and the second half with 0, the test 1 would have passed although the sequence is highly non-random.
(2) Through this test it is intended to ensure that frequencies of 1 and 0 are evenly distributed across the entire n-bit sequence.

**Background information in respect of randomness**:

(1) The n-bit string is divided in non-overlapping N blocks each of M-bit, where N=floor of [n/M]. The $(N+1)^{th}$ block having bits less than M is neglected. M should be taken as reasonably small.
(2) If the proportion of 1s in each block is approximately ½, the n-bit string can be termed as random.
(3) Proportion $\Pi_i$ of 1s in each block is given by,

$$\Pi_i = ( \sum \varepsilon_{(i-1)M+j} ) / M, \text{ for the summation j varies from 1 to M}.$$

where i varies from 1 to N.

(4) Chi-square ($\chi^2$) is given by,

$$\chi^2 = 4M \sum (\Pi_i - \tfrac{1}{2})^2, \text{ for } i = 1 \text{ to } N.$$

**Focus of Computation**:

(1) For each block, $\Pi_i$ as given above is calculated for i varying from 1 to N.
(2) The chi-square ($\chi^2$) as given above is computed.
(3) N is the degrees of freedom.
(4) $a = \frac{1}{2}(N)$ and $x = \frac{1}{2}(\chi^2)$
(5) P-value = $1 - \Gamma(a, x) / \Gamma(a, \infty)$.

### 3.3. Test 3: Runs Test (1 P-value with 1 parameter, x)

**Prelude**:
(1) Runs of length k means exactly k identical bits bounded by bits of opposite value.
(2) Through this test it is intended to see if the frequencies of runs of 1s and 0s of various lengths would be within limits of randomness.

**Background information in respect of randomness**:

(1) $\varepsilon_i$ = bit value at $i^{th}$ position of the n-bit string ; $\Pi = (\sum \varepsilon_i )/n$ , $\sum$ is for i varying from 1 to n.
(2) A checking parameter ($\tau$) is defined as, $\tau = 2/\sqrt{n}$.
(3) If $|\Pi - \frac{1}{2}| \geq \tau$, it is not necessary to run the present test, since the Test 1 would fail for the sequence. When $|\Pi - \frac{1}{2}| < \tau$, the Runs test is executed.

**Focus of Computation**:

(1) $V_n$ (obs) = $\sum r(k) + 1$, and $r(k)=0$ if $\varepsilon_k = \varepsilon_{k+1}$ and $r(k)=1$ otherwise. Bit 1 is to be added at the end.
(2) $x = | V_n (obs) - 2n \Pi(1- \Pi) | / [2\sqrt{(2n)} \Pi (1- \Pi)]$
(3) P-value = $1 - erf(x)$.

### 3.4. Test 4: Longest Run Test of 1s in a Block (1 P-value with 2 parameters, a & x)

**Prelude**:

(1) Through this test it is intended to see if the frequencies of longest run of 1s of various lengths appearing in the sequence are consistent with that expected for a random sequence.

(2) To execute the test the n-bit string is divided in N non-overlapping blocks each of M-bit such that N= floor of [n/M] as being done for Test 2. The additional bits are neglected.

**Background information in respect of randomness**:

(1) Considering all blocks, $\nu_i$ represents sum of all frequencies of longest runs of particular 1s appearing in each block.
(2) For the sake computation $\nu_i$ is divided into (K+1) classes with i varying from 0 to K. Among n, M, N and K, an empirical relation as given below is proposed.

| Minimum n | Minimum M | K | Minimum N |
|---|---|---|---|
| 128 | 8 | 3 | 16 |
| 6272 | 128 | 5 | 49 |
| 750000 | $10^4$ | 6 | 75 |

(3) $v_0$ is the number of blocks where '1' is the longest run of 1s or all 0s in the block, $v_1$ is the number of blocks where '11' is the longest run of 1s, $v_2$ is the same for '111', so on so forth. For K=3, $v_3$ is the number of blocks where the longest runs of 1s is '1111' or more.
(4) The number of times the longest runs of 1s are appearing in a particular block is not considered.
(5) Considering randomness the theoretical studies on probabilities of occurrences of longest runs of 1s have been undertaken in detail for M=8 & K=3, M=128 & K=5, M=512 & K=5, M=1000 & K=5 and M=10000 & K=6. A representative set of one such values for M=8 & K=3 are given below,

| | 4 classes | Probabilities |
|---|---|---|
| | $v_0 \leq$ '1', one 1s or no 1s | 0.2148 |
| M = 8 & K = 3 | $v_1$ = '11' (2 ones) | 0.3672 |
| | $v_2$ = '111' (3 ones) | 0.2305 |
| | $v_3 \geq$ '1111' (4 ones or more) | 0.1875 |

(6) The probabilities of occurrences of longest runs of 5 ones or more are so small that these are clubbed together in $v_3$.
(7) It may be noted for M=128 & K=5, the 6 classes are marked as $v_0 \leq 4$ ones, $v_1 = 5$ ones, $v_2 = 6$ ones, $v_3 = 7$ ones, $v_4 = 8$ ones and $v_5 \geq 9$ ones. Other cluster of classes, e.g. M=512 & K=5, M=1000 & K=5 and M=10000 & K=6, have their respective classes with probabilities. All related data are well compiled in the NIST document
(8) It may be noted that $\sum v_i = N$ and the sum of probabilities for a particular (M, K) group is unity.

**Focus of Computation**:

(1) n-bit string is divided in N blocks each of 8-bit long.
(2) The longest of runs of 1s are observed in each block and the appropriate classes ($v_i$) are incremented. And at the end of the $N^{th}$ block all the classes appropriately filled. It may noted that the summation $v_i$ classes for i = 0 to k becomes N.
(3) Chi-square ($\chi^2$) statistic is given by,
   $\chi^2 = \sum (v_i - N\pi_i)^2 / N\pi_i$, where i varies from 0 to k.
(4) a = ½ (K)   and   x = ½ ($\chi^2$)
(5) P-value = 1 − $\Gamma$(a, x) / $\Gamma$(a, ∞)

**3.5. Test 5: Binary Matrix Rank Test (1 P-value with 2 parameters, a & x)**

**Prelude**:

(1) Through this test it is intended to see if the n-bit string has repetitive patterns across its entire sequence. The n-bit string is sequentially divided into N disjoint blocks and it is endeavored to see linear dependence among its fixed length substrings of each block.
(2) Each block is represented by a matrix of M rows and Q columns such that N= floor of (n/MQ). The remaining unused bits are discarded. Usually both M and Q are taken as 32.

(3) Through the test it is intended to calculate the rank of all sub-matrices. For a sub-matrix $M_1$ of order M, the search is for its rank. If its determinant is non-zero, its rank is M. If its determinant is zero, the determinants of all its sub-matrices of order (M-1) are calculated. If at least one determinants of order (M-1) is non-zero, the rank of $M_1$ is M-1. This is the way one has to go lower order matrices to find its rank.

(4) For a full rank sub-matrix, one can conclude that it has no repetitive patterns.

**Background information in respect of randomness**:

(1) There were lots of theoretical studies related to rank of square binary matrix of order (M=10) and above. The study indicates that the probability ($P_M$) of the rank of a square binary matrix of orders M, M-1 and M-2 is zero are given as follows,

$P_M = \Pi [ 1 – 2^{-j} ]$ , for j varying from 1 to ∞
$= 0.5 * 0.75 * 0.875 * 0.9375 * 0.9843755 * 0.9921875 * 0.996009375 * .....$
$= 0.2888......$

$P_{M-1} = 2 P_M = 0.5776.......$
$P_{M-2} = (4/9) P_M = 0.1284....$

and all other probabilities are very small ($\leq 0.005$).

(2) Considering the very small probability values of $P_{M-2}$, $P_{M-3}$ etc., it is assumed that the matrices of order M-2 and less can be clubbed with $P_{M-2}$ and $P_{M-2}$ will then assume a value of 0.1336 instead of 0.1284. Please note that 0.2888 + 0.5776 + 0.1336 = 1.

(3) From probability consideration, the degrees of freedom (K) will 2, since there are 3 classes.

(4) For the sake of convenience of computation, M is taken as 32 and each sub-matrix would contain 1024 bits (=32 * 32).

**Focus of Computation**:

(1) The determinants of all the sub-matrices of order 32 is determined and non-zero ones are counted.
(2) $F_M$ = Number of sub-matrices having full rank M.
(3) $F_{M-1}$ = Number of sub-matrices with rank (M-1) from among the (N - $F_M$) sub-matrices.
(4) $F_{M-2}$ = N - $F_M$ - $F_{M-1}$ = Number of sub-matrices with rank (M-2) and less.
(5) $\chi^2 = \sum (F_i – N P_i)^2 / N P_i$ , with i = M, M-1 and M-2.
(6) a = ½ (K)   and   x = ½ ($\chi^2$)
(7) For a = 1, $\Gamma$ (a, x) = [1 – exp(-x)] and $\Gamma$ (a, ∞) = 1.
(8) P-value = 1 – $\Gamma$ (a, x) / $\Gamma$ (a, ∞) = exp(-x).

**3.6. Test 6: Discrete Fourier Transform Test (1 P-value with 1 parameter, x)**

**Prelude**:

(1) Through this test it is intended to see if the n-bit string has periodic features across its entire sequence. By periodic features one understands repetitive patterns that are close to each other.

(2) The focus of the test is to undertake Discrete Fourier Transform (DFT) of each bit of the sequence and to ascertain their peak heights.
(3) Considering randomness one can find a peak height threshold value (T). If at most 5% of the peak heights are more than T, the sequence can be termed as random.

**Background information in respect of randomness**:

(1) DFT produces a sequence of complex variables to represent periodic components of different frequencies. The DFT component of the $j^{th}$ bit is given by $F_j$ as,

$F_j = \sum [x_k \exp(j * 2\pi i(k-1) / n)]$     for k=1,2,…..,n and $x_k$ is the $k^{th}$ bit.

(2) The peak height threshold value (T) is calculated using the relation $T=\sqrt{[(\log(1/0.05))*n]}$.
(3) Because of the symmetry of the real to complex-value transform, only values of j are considered from 0 to (n/2 – 1) instead n.

**Focus of Computation**:

(1) Each bit of 0 and 1 in the n-bit sequence is represented by -1 and 1 respectively by using a relation $X_i = 2\varepsilon_i - 1$, where $X_i$ represents new value of the bit $\varepsilon_i$ at the $i^{th}$ position. (i varies from 0 to n).
(2) T is calculated using the relation stated above.
(3) $N_0$ = Expected theoretical (95%) number of peaks under the assumption of randomness = 0.95n/2.
(4) Following the expression given above, magnitude (M) of $F_j$ is calculated for j = 0 to (n/2 – 1).
(5) $N_1$ = Number of peaks in M that are less than T.
(6) $d = (N_1-N_0)/\sqrt{[n(0.95)(0.05)/2]}$.
(7) $x = |d|/\sqrt{2}$.
(8) P-value = 1 – erf (x).

### 3.7. Test 7: Non-overlapping Template Matching Test (1 P-value with 2 parameters: a, x)

**Prelude**:

(1) Through this test one intends to detect template matching in a non-overlapping manner, i.e. it looks for occurrences of pre-specified bit-string and to see if the numbers of such occurrences are within the statistical limit of a sequence under the assumption of randomness.
(2) An m-bit window is considered to search for specific m-bit pattern. If the pattern is not found, the window slides one bit position. If the pattern is found, the window is reset to the bit next to the found pattern.
(3) This test detects generators producing too many occurrences of non-periodic patterns (aperiodic).

**Background information in respect of randomness**:

(1) For random sequences the Central limit theorem is assumed to be applicable.

(2) Mean (μ) and variance (σ²) are calculated based on approximate normal distribution are given by,

$$\mu = (M-m+1)/2^m \text{ and } \sigma^2 = M[1/2^m - (2m-1)/2^{2m}],$$

where m is the fixed length of the non-periodic pattern appearing M times.

**Focus of Computation**:

(1) n-bit sequence is divided in N non-overlapping blocks each of M-bit where N= floor of [n/M]. The unused bits are discarded.
(2) Mean μ and variance σ² are calculated following the expression given above..
(3) $W_j$ = Number of times the specified pattern is found in the j$^{th}$ block. The matching search is continued for all blocks of j from 1 to N.
(4) $\chi^2 = \sum (W_j - \mu)^2 / \sigma^2$, for j =1 to N
(5) a = ½ (N) and x = ½ ($\chi^2$)
(6) P-value = P-value = 1 − $\Gamma$(a, x) / $\Gamma$(a, ∞)

## 3.8. Test 8: Overlapping Template Matching Test (1 P-value with 2 parameters, a & x)

**Prelude**:

(1) Through this test one intends to detect template matching in an overlapping manner, i.e. it looks for occurrences of pre-specified bit-string and to see if the number of such occurrences as against a sequence under the assumption of randomness.
(2) An m-bit window is considered to search for specific m-bit pattern. The window always slides one bit position next, whether the pattern is found or not.
(3) For this test the Poisson asymptotic distribution is assumed to be followed.
(4) n-bit string is divided in N blocks each of M-bit such that N= floor of [n/M]. Extra bits are discarded.
(5) Six classes ($v_i$) are considered with i = 0 to 5. The explanation of $v_i$ follows. $v_0$ =329 : m-bit pattern is not found in 329 blocks; $v_1$ =164: m-bit pattern is found once in 164 blocks; $v_2$ =150: m-bit pattern is found twice in 150 blocks; $v_3$ =111: m-bit pattern is found thrice in 111 blocks; $v_4$ =78: m-bit pattern is found four times in 78 blocks; $v_5$ =136: m-bit pattern is found five times or more in 136 blocks.
(6) Degrees of freedom are K, for this test it is 5.
(7) The test detects any irregular occurrences of any periodic pattern.
(8) The test sometime rejects sequences with too many or too few occurrences of m-runs of ones.

**Background information in respect of randomness**:

(1) There were many theoretical studies in respect of overlapping template matching. For computing theoretical probabilities ($\pi_i$) corresponding to classes $v_i$, values of λ and η are calculated as,

$$\lambda = (M-m+1)/2^m \text{ and } \eta = \lambda/2$$

where m is the fixed length of the non-periodic pattern and M is the bit size of each block..

(2) Under the assumption of randomness the theoretical probability values are available in standard literatures as follows,

$\pi_0 = 0.324652$, $\quad \pi_1 = 0.182617$, $\quad \pi_2 = 0.142670$, $\quad \pi_3 = 0.106645$,
$\pi_4 = 0.077147$, $\quad \pi_5 = 0.166269$.

It may be noted that $\lambda$ and $\eta$ are necessary to calculate all values of $\pi_i$.

**Focus of Computation**:

(1) Few recommendations: (i) $n \geq 10^6$, (ii) m = 9 or 10, (iii) $N > 5/\min(\pi_i)$, (iv) $n \geq MN$, (v) $\lambda \approx 2$, (vi) $m \approx \log_2 M$, (vii) $K \approx 2\lambda$, (viii) The $\pi_i$ values given are exclusively for K=5.
(2) The overlapping count of the m-bit window is undertaken for all N blocks and the array of $v_i$ classes are correspondingly filled.
(3) $\chi^2 = \sum(v_i - N\pi_i)^2 / N\pi_i$, for i = 0 to 5.
(4) $a = \frac{1}{2}(K)$ and $x = \frac{1}{2}(\chi^2)$
(5) P-value = $1 - \Gamma(a, x) / \Gamma(a, \infty)$.

**3.9. Test 9: Maurer's "Universal Statistical" Test (1 P-value with 1 parameter, x)**

**Prelude**:

(1) The test focuses to measure distances in terms L-bit block-numbers between L-bit matching patterns. The sum of $\log_2$ distances between L-bit matching patterns is necessary for statistic distribution.
(2) Through this test one will be in a position to conclude whether the sequence could be significantly compressed or not. A significantly compressible sequence is considered to be non-random.
(3) Standard normal (Gaussian) density distribution is used to obtain expected value of the test statistic function ($f_n$) along with its standard deviation ($\sigma$) under the assumption of randomness.
(4) n-bit string is divided into two blocks: one is the initialization segment with Q number of L-bit blocks and another is the test segment with K number of L-bit blocks. Unused bits are discarded.

**Background information in respect of randomness**:

(1) The test looks back through the entire sequence while walking through the test segment consisting of K number of L-bit blocks, checking for a match with nearest previous exact L-bit template and recording the distance – in number of blocks – to that previous match. The algorithm computes the $\log_2$ of all such distances for all the L-bit templates in the test segment (giving, effectively, the number of digits in the binary expansion of each distance). Then it averages over all the expansion lengths by the number of K test blocks as given below,

$\quad f_n = (1/K) [\sum \log_2 (\text{\# indices since previous occurrence of } I^{th} \text{ template})]$

where for the summation I varies from (Q+1) to (Q+K). Based on standard normal (Gaussian) density distribution the expected value of the theoretical test statistic $E(f_n)$ is derived as,

$$E(f_n) = 2^{-L} \sum (1 - 2^{-L})^{I-1} \cdot \log_2 I, \text{ where I varies from 1 to } \infty$$

A separate expression for Variance (L) is also given. The variance is related to the theoretical standard deviation ($\sigma$) as,

$$\sigma = C \sqrt{[\text{variance (L) / K}]},$$

where $C = 0.7 - (0.8/L) + (4 + 32/L)(K^{-3/L}/15)$.

A dynamic look-up table has been generated making use of the integer representation of the binary bits constituting the L-bit template blocks of different sizes. The look-up table for L varying from 6 to 16 is given below,

| L | Expected Value $E(f_n)$ | Variance | L | Expected Value $E(f_n)$ | Variance |
|---|---|---|---|---|---|
| 6 | 5.2177052 | 2.954 | 12 | 11.168765 | 3.401 |
| 7 | 6.1962507 | 3.125 | 13 | 12.168070 | 3.410 |
| 8 | 7.1836656 | 3.238 | 14 | 13.167693 | 3.416 |
| 9 | 8.1764248 | 3.311 | 15 | 14.167488 | 3.419 |
| 10 | 9.1723243 | 3.356 | 16 | 15.167379 | 3.421 |
| 11 | 10.170032 | 3.384 | | | |

**Focus of Computation**:

(1) A table with possible L-bit value is created where last occurrence of the block number of each L-bit is noted. In the Test segment K, each block is checked and the distance between present block and the block where same L-bit block occurs last time is calculated. The previous block number is replaced by the current block number.
(2) Test statistic function ($f_n$) is calculated based on the following expression,

$$f_n = 1/K \sum \log_2 (i - T_j), \text{ for } i = Q+1 \text{ to } Q+K$$

where j is the decimal representation of the content of the $i^{th}$ L-bit block and $T_j$ is the table entry. The previous of table entry of $T_j$ is replaced by the current $i^{th}$ block number.
(3) The standard deviation ($\sigma$) is computed based on the expression given above and the corresponding value of variance given in the Table above.
(4) $x = |f_n - E(f_n)| / \sqrt{2}\sigma$
(5) P-value = $1 - \text{erf}(x)$.

### 3.10. Test 10: Linear Complexity Test (1 P-value with 2 parameters, a and x)

**Prelude**:

(1) A long bit string is usually obtained from a LFSR (Linear Feedback Shift Register).
(2) The bit sequence from which a longer LFSR is obtained can be termed as random, while the shorter LFSR indicates non-randomness.

(3) The Berlekamp-Massey Algorithm is adopted to obtain a LFSR.
(4) The linear complexity test looks for length of LFSR and determines if the bit sequence from which the LFSR is obtained is random or not.

**Background information in respect of randomness**:

(1) A long n-bit sequence is divided into N blocks, each of M-bit.
(2) Considering randomness the mean length of LFSR ($\mu$) of M-bit string is given by,

$$\mu = (M/2) + (9 + (-1)^{M+1})/36 - [(M/3) + (2/9)]/2^M$$

(3) The statistical deviation ($T_i$) of a LFSR of length ($L_i$) is given by,

$$T_i = (-1)^M (L_i - \mu) + 2/9$$

(4) Depending on values of $T_i$, N blocks are divided in 7 fixed groups ($v_i$) where i varies from 0 to 6, based on the following considerations:

$v_0$ ( $T_i \leq -2.5$), $v_1$ ( $-2.5 < T_i \leq -1.5$), $v_2$ ( $-1.5 < T_i \leq -0.5$), $v_3$ ( $-0.5 < T_i \leq +0.5$),

$v_4$ ($+0.5 < T_i \leq +1.5$), $v_5$ ($+1.5 < T_i \leq +2.5$), $v_6$ ($T_i \geq +2.5$),.

(5) The theoretical probabilities ($\pi_i$) of each of the 7 groups stated above are obtained from standard literature as,

$\pi_0 = 0.01047$, $\pi_1 = 0.03125$, $\pi_2 = 0.125$, $\pi_3 = 0.5$,

$\pi_4 = 0.25$, $\pi_5 = 0.0625$, $\pi_6 = 0.02078$.

**Focus of Computation**:

(1) The focus of the test is to find LFSR for each M-bit sub-stings and to find its length ($L_i$).
(2) $\mu$ is calculated for the value of M.
(3) $T_i$ is calculated for each of N blocks. Depending on the value of $T_i$ the appropriate $v_i$ array is incremented. One may note that the sum of $v_i$ for i = 0 to 6 is N.
(4) Here the degrees of freedom (K) are considered to be 6.
(5) Had it been that there was no group; T would be in one group. Creation of 7 groups provides T a choice of additional 6 groups – hence degrees of freedom are 6.
(6) The chi-square statistic ($\chi^2$) is calculated as,

$$\chi^2 = \sum [(v_i - N \pi_i)^2 / N\pi_i], \text{ for } i = 0 \text{ to } K$$

(7) a = ½ (K) and x = ½ ($\chi^2$)
(8) P-value = 1 – $\Gamma$ (a, x) / $\Gamma$ (a, ∞)

### 3.11. Test 11: Serial Test (2 P-values each one with 2 parameters, a & x)

**Prelude:**

(1) In a long n-bit random sequence, every m-bit pattern has the same chance of appearing as every other m-bit patterns.
(2) The number of occurrences of the $2^m$ m-bit overlapping patterns is approximately the same as would be expected of a random sequence.

(3) In n-bit sequence, each of all m-bit patterns is expected to occur $A_m$ times, where $A_m = n/2^m$.
(4) The serial test counts the frequency of all possible overlapping m-bit patterns across the entire n-bit sequence and based on the deviations of each of all counts together one intends to see if the sequence can be termed as random or not.

**Background information in respect of randomness**:
(1) Let $v_i$ represents frequency counts for i varying from 0 to $2^m$ where i denotes the decimal value of a particular m-bit pattern.
(2) The psi-square statistic ($\psi_m^2$), similar to chi-square ($\chi^2$), is given by,

$$\psi_m^2 = \sum (v_i - A_m P)^2 / A_m, \text{ summation is from } i = 0 \text{ to } (2^m - 1)$$
$$= (1/A_m) \sum v_i^2 - n$$

(3) The chi-square statistic ($\chi^2$) in the present case is,

$$\Delta \psi_m^2 = \psi_m^2 - \psi_{m-1}^2$$
$$\Delta \psi_{m-1}^2 = \psi_{m-1}^2 - \psi_{m-2}^2$$
$$\Delta^2 \psi_m^2 = \Delta \psi_m^2 - \Delta \psi_{m-1}^2 = \psi_m^2 - 2\psi_{m-1}^2 + \psi_{m-2}^2$$

(4) Here the $\Delta \psi_m^2$ is the $\chi^2$ distribution with $K_1 = 2^{m-1}$ degrees of freedom and $\Delta^2 \psi_m^2$ is another $\chi^2$ distribution with $K_2 = 2^{m-2}$ degrees of freedom.
(5) Two chi-square ($\chi^2$) distributions coupled with two degrees of freedom gives rise two P-values.
(6) Value of m is usually small and $m \leq \text{floor} [\log_2 (n)] - 2$.
(7) Serial Test turns out to be the frequency test (Test 1) if m=1.

**Focus of Computation**:
(1) For m-bit pattern $v_i$ is counted for $i = 0$ to $(2^m - 1)$; $\psi_m^2$ is computed with $A_m = n/2^m$.
(2) For (m-1)-bit pattern $v_i$ is counted for $i = 0$ to $(2^{m-1} - 1)$; $\psi_{m-1}^2$ is computed with $A_{m-1} = n/2^{m-1}$.
(3) For (m-2)-bit pattern $v_i$ is counted for $i = 0$ to $(2^{m-2} - 1)$; $\psi_{m-2}^2$ is computed with $A_{m-2} = n/2^{m-2}$.
(4) Based on $\psi_m^2$ and $\psi_{m-1}^2$, $\Delta \psi_m^2$ is computed and based $\psi_m^2$, $\psi_{m-1}^2$ and $\psi_{m-2}^2$, $\Delta^2 \psi_m^2$ is computed.
(5) Considering $a_1 = \frac{1}{2} (K_1)$ and $x_1 = \frac{1}{2} (\Delta \psi_m^2)$,

$$\text{P-value}_1 = 1 - \Gamma(a_1, x_1) / \Gamma(a_1, \infty)$$

(6) Considering $a_2 = \frac{1}{2} (K_2)$ and $x_2 = \frac{1}{2} (\Delta^2 \psi_m^2)$,

$$\text{P-value}_2 = 1 - \Gamma(a_2, x_2) / \Gamma(a_2, \infty)$$

### 3.12. Test 12: Approximate Entropy Test (1 P-value with 2 parameters, a & x)

**Prelude:**

(1) Entropy is a test of randomness based on repeating patterns. Larger is the entropy larger is the randomness.
(2) For n-bit string the entropy is measured by comparing the frequency of overlapping patterns of all possible m-bit patterns with that of (m+1)-bit patterns. The comparison between entropies of m-bit and (m+1)-bit patterns is termed as approximate entropy, *ApEn* (m), which is compared against the expected result of a random sequence.
(3) For a random sequence, the *ApEn* (m) is a maximum value projected as *ln2*.
(4) Test of the binary sequence of e, $\pi$, $\sqrt{2}$ and $\sqrt{3}$ has shown that $\sqrt{3}$ is more irregular than $\pi$ and their values show a limiting convergence towards *ln2*.

**Background information in respect of randomness**:

(1) For counting m-bit matching patterns, (m-1) bits taken from the beginning of the sequence are appended at the end of the given n-bit string in the form.
(2) Let $v_i$ represents overlapping frequency counts of a particular m-bit pattern for i varying from 0 to $2^m$, where i denotes the decimal value of a particular m-bit pattern.
(3) $C_i^m = v_i / n$, $\pi_i = C_i^m$ and $\Phi^m = \sum (\pi_i \ln \pi_i)$, for i = 0 to ($2^m$ -1)
(4) For counting (m+1)-bit matching patterns, first m bits are appended at the end of the given n-bit string.
(5) Similarly $v_i$ represents overlapping frequency counts of a particular (m+1)-bit pattern for i varying from 0 to $2^{m+1}$, where i denotes the decimal value of a particular (m+1)-bit pattern.
(6) $C_i^{m+1} = v_i / n$, $\pi_i = C_i^{m+1}$ and $\Phi^{m+1} = \sum (\pi_i \ln \pi_i)$, for i = 0 to ($2^{m+1}$ -1)
(7) $ApEn (m) = \Phi^m - \Phi^{m+1}$
(8) K = Degrees of freedom = $2^m$
(9) $\chi^2 = 2 n [\ln 2 - ApEn (m)]$

**Focus of Computation**:

(1) For counting m-bit matching patterns, $v_i$ is counted in a overlapping manner across the appended (n+m-1)-bit sequence for all possible m-bit patterns, i varies from 0 to ($2^m$ -1).
(2) Based on $2^m$ types of $v_i$, values of $C_i^m$, $\pi_i$ and $\Phi^m$ are computed.
(3) Again for counting (m+1)-bit matching patterns, $v_i$ is counted in a overlapping manner across the appended (n+m)-bit sequence for all possible (m+1)-bit patterns, i varies from 0 to ($2^{m+1}$ -1).
(4) Based on $2^{m+1}$ types of $v_i$, values of $C_i^{m+1}$, $\pi_i$ and $\Phi^{m+1}$ are computed.
(5) $\chi^2 = 2 n [\ln 2 - ApEn (m)]$ where ApEn (m) = $\Phi^m - \Phi^{m+1}$
(6) Now a = ½ (K) and x = ½ ($\chi^2$).
(7) P-value = 1 $- \Gamma (a, x) / \Gamma (a, \infty)$

### 3.13. Test 13: Cumulative Sums Test (2 P-values each one with 1 parameter, x)

**Prelude:**

(1) The cumulative sums test looks whether 1s or 0s are occurring in large numbers at early stages or at later stages or 1s and 0s are intermixed evenly across the entire sequence.

**Background information in respect of randomness:**

(1) Since the distribution of cumulative sums is being looked into, the P-value is calculated following the Normal distribution function ($\Phi$) given by,

$$\Phi(z) = (1/\sqrt{(2\pi)}) \int \exp(-u^2/2)\,du ,$$

where lower and upper limits of u are $-\infty$ and z respectively.

**Focus of Computation:**

(1) Across the entire n-bit sequence, the 0s are made -1 as it is done in Test 1. The cumulative sums of adjusted (-1, +1) of $X_i$ digits of the sequence is obtained as $S_i = S_{i-1} + X_i$ with i = 1 to n and $S_0 = 0$. The cumulative sums may be considered as Random Walk.
(2) In Test 1 the sum was of adjusted (-1, +1) of all $X_i$ digits and it was seen if the summation falls within the accepted range of randomness.
(3) Here the maximum magnitude value of the cumulative sums $S_i$ ($S_i^{max} = m$) is being looked into. Based on m, z is calculated as ($m/\sqrt{n}$). If z is large, the bit sequence is considered to be non-random.
(4) The cumulative sums can undertaken in a forward manner, i.e. from start to end (termed as Mode 0) and also in a backward manner, i.e. from end to start (termed as Mode 1). For each of the two cases, two sets of m (= $S_i^{max}$) and z are noted.
(5) The P-value is computed using the following equation involving Normal distribution function ($\Phi$).

**P-value = 1 – $\sum [\Phi\{(4k+1)z\} – \Phi\{(4k-1)z\}] + \sum [\Phi\{(4k+3)z\} – \Phi\{(4k+1)z\}]$**

For the 1st $\sum$, k = (-n/z +1)/4 to (n/z – 1)/4, while for the 2nd $\sum$, k = (-n/z – 3)/4 to (n/z – 1)/4.

(6) Two P-values are calculated following the Normal distribution function – one corresponding to forward cumulative sums and the other, to backward cumulative sums.

### 3.14. Test 14: Random Excursions Test (8 P-values each one with 2 parameters a & x)

**Prelude:**

(1) The Random Excursions Test intends to look if the number of visits to a particular cumulative sums state within a cycle falls into a category that is expected of random sequence.
(2) Eight states, e.g. -4 , -3 , -2 , -1 and +1 , +2 , +3 , +4 are looked into – visits to states greater than +4 are clubbed within the visits to +4 state and visits to states lesser then -4 are clubbed within the visits to -4 state.

**Background information in respect of randomness:**

(1) $\pi_k(s)$ is defined as the theoretical probability of k number of visits to a state s. Expressions of $\pi_k(s)$ for k = 0,1,2,3,4 and 5 are being theoretically derived.
(2) $\pi_0(s)$ = Probability of 0 number of visits to a state s = $1 - 1/(2|s|)$.
(3) $\pi_k(s)$ = Probability of k number of visits to a state s = $1/4s^2 (1 - 1/(2|s|))^{(k-1)}$, k=1, 2, 3 and 4.
(4) $\pi_5(s)$ = Probability of 5 number of visits or more to a state s = $1/(2|s|)(1 - 1/(2|s|))^4$.
(5) The fourteen states, e.g. ±1, ±2, ±3, ±4, ±5, ±6, ±7 have been theoretically considered. The study indicates that the states ±5, ±6 and ±7 have very low probability occurrences and this is the reason that first eight states are considered for practical situation. The theoretical probability values are shown in the NIST document and are shown here for ready reference.

|        | $\pi_0(s)$ | $\pi_1(s)$ | $\pi_2(s)$ | $\pi_3(s)$ | $\pi_4(s)$ | $\pi_5(s)$ |
|--------|--------|--------|--------|--------|--------|--------|
| s = ±1 | 0.5000 | 0.2500 | 0.1250 | 0.0625 | 0.0312 | 0.0312 |
| s = ±2 | 0.7500 | 0.0625 | 0.0469 | 0.0352 | 0.0264 | 0.0791 |
| s = ±3 | 0.8333 | 0.0278 | 0.0231 | 0.0193 | 0.0161 | 0.0804 |
| s = ±4 | 0.8750 | 0.0156 | 0.0137 | 0.0120 | 0.0105 | 0.0733 |
| s = ±5 | 0.9000 | 0.0100 | 0.0090 | 0.0081 | 0.0073 | 0.0656 |
| s = ±6 | 0.9167 | 0.0069 | 0.0064 | 0.0058 | 0.0053 | 0.0588 |
| s = ±7 | 0.9286 | 0.0051 | 0.0047 | 0.0044 | 0.0041 | 0.0531 |

(6) It may be noted that $\sum \pi_k(s)$ for K = 0 to 5 is unity for a visit to a particular state x.

**Focus of Computation:**

(1) Across the entire n-bit sequence, the 0s are made -1 as it is done in Test 1. The cumulative sums of adjusted (-1, +1) of $X_i$ digits of the sequence is obtained as $S_i = S_{i-1} + X_i$ with i = 1 to n and $S_0 = 0$ and also $S_{n+1} = 0$
(2) If the cumulative sums crosses zero (J-1) times, J is termed as the number cycles considering the zero crossing point at $S_{n+1}$.
(3) In the event J is too small, the sequence is considered to be non-random. A 10 Lac bit sequence is considered non-random if J is less than 500.
(4) $v_k(s)$ = Frequency of k-times of visit to the state x during J excursions. For the sake of computation one can consider,

$$v_k(s) = \sum v_k^j(s), \text{ summation is taken for j=1 to J}$$

If the number of visits to the state s during $j^{th}$ excursion is exactly equal to k, then $v_k^j(s) = 1$ else $v_k^j(s) = 0$.
(5) For $v_k(s)$ with k ≥ 5, the count data is being put in $v_5(s)$.
(6) For each state, the chi-square statistic is calculated as,

$$\chi^2 = \sum [v_k(s) - J\pi_k(s)]^2 / J\pi_k(s), \text{ with K = 0 to 5.}$$

(7) $a = \frac{1}{2}(K)$ and $x = \frac{1}{2}(\chi^2)$

(8) P-value = $1 - \Gamma(a, x) / \Gamma(a, \infty)$

(9) There are eight states – hence there will be eight P-values corresponding to each state.

### 3.15. Test 15: Random Excursions Variant Test (18 P-values each one with parameter, x)

**Prelude:**

(1) The Random Excursions Variant test looks for number of visits to a particular state in cumulative sums of random walk across the entire bit sequence and estimates deviations from expected number of visits in the random walk considering randomness.

(2) 18 states, e.g. s = –9, –8, –7, –6, –5, –4, –3, –2, –1, +1, +2, +3, +4, +5, +6, +7, +8, +9 are considered.

**Background information in respect of randomness**:

(1) Statistic Variation ($\sigma$) = $2(2|s| - 1)$ in respect of random walk for visit to different states.

**Focus of Computation**:

(1) Across the entire n-bit sequence, the 0s are made -1 as it is done in Test 1. The cumulative sums of adjusted (-1, +1) of $X_i$ digits of the sequence is obtained as $S_i = S_{i-1} + X_i$ with i = 1 to n and $S_0 = 0$ and also $S_{n+1} = 0$

(2) If the cumulative sums crosses zero (J-1) times, J is termed as the number cycles considering the zero crossing point at $S_{n+1}$.

(3) $\xi(s)$ is defined as the total number of times that a state **s** is visited across all J cycles.

(4) Here $x = |\xi(s) - J| / \sqrt{2 J \sigma}$.

(5) P-value = $1 - \text{erf}(x)$.

There are eighteen states – hence eighteen P-values corresponding to each state are calculated.

### 4. Results and Discussion:

The package containing 15 tests is used to test strength of randomness of three PRNG algorithms, namely two Linear Congruent Generators (LCG) by Park & Miller and Knuth and the Blum-Blum-Shub Generator (BBSG). For each algorithm more than $10^6$ bits are taken for one sequence and such 300 sequences are used. The sequence length for individual tests varies from test to test. The lengths of bits required for each test are taken from one sequence from the beginning and such methodology is adopted for 300 sequences. The lengths of bit-sequences taken for each test are given in Table A. The 15 tests are executed on these 300 sequences. A test is unsuccessful when P-value < 0.01 and then the sequence under test should be considered as non-random.

Table A: Lengths of bit-sequence for different tests

| Test No. | Test Name | Required length of bit-sequence (n) | Used length of bit-sequence in this package |
|---|---|---|---|
| 1 | Frequency (Monobit) Test | n ≥ 100 | 100 |
| 2 | Frequency Test within a Block | n ≥ 9,000 | 9000 |
| 3 | Runs Test | n ≥ 100 | 100 |
| 4 | Test for the Longest Run of Ones in a Block | n ≥ 128 / 6,272 / 7,50,000 | 128 |
| 5 | Binary Matrix Rank Test | n ≥ 38,912 | 38912 |
| 6 | Discrete Fourier Transform (Spectral) Test | n ≥ 1,000 | 1000 |
| 7 | Non-overlapping Template Matching Test | n ≥ 10,48,576 | 1048576 |
| 8 | Overlapping Template Matching Test | n ≥ 10,00,000 | 1000000 |
| 9 | Maurer's "Universal Statistical" Test | n ≥ 13,42,400 | 1342400 |
| 10 | Linear Complexity Test | n ≥ 10,00,000 | 1000000 |
| 11 | Serial Test | n ≥ 10,00,000 | 1000000 |
| 12 | Approximate Entropy Test | n = 100 to 1,000 | 100 |
| 13 | Cumulative Sums (Cusum) Test | n ≥ 100 | 100 |
| 14 | Random Excursions Test | n ≥ 10,00,000 | 1000000 |
| 15 | Random Excursions Variant Test | n ≥ 10,00,000 | 1000000 |

**Proportion of passing a test based on P-values:**

For considering the proportion of passing of a test, it is necessary to consider large number of samples of bit-sequences generated by a PRNG. If m samples of bit sequences obtained from one PRNG algorithm are tested by a test producing one P-value, then a statistical threshold value is defined as,

$$\text{Threshold value (T-value)} = (1-\alpha) - 3\sqrt{[\alpha(1-\alpha)/m]}$$

where significance level ($\alpha$) = 0.01. The size of m should greater than inverse of $\alpha$. If m=300, T-value = 0.972766. This means that such a test is considered statistically successful, if at least 292 sequences out of the given 300 sequences do pass the test.

For identical experiments for a test producing 'n' P-values, for the calculation of T-value one should consider (m*n) instead of m. With same values of $\alpha$ and m, the T-value is 0.983907 for n=8 (tests 14). Such a test is considered statistically successful if at least 296 sequences out of the given 300 sequences do pass the test.

The results for LCG by Knuth, BBSG and LCG by Park & Miller are shown in Appendices I, II and III respectively. Graphical representation for proportion of passing of BBSG is given below

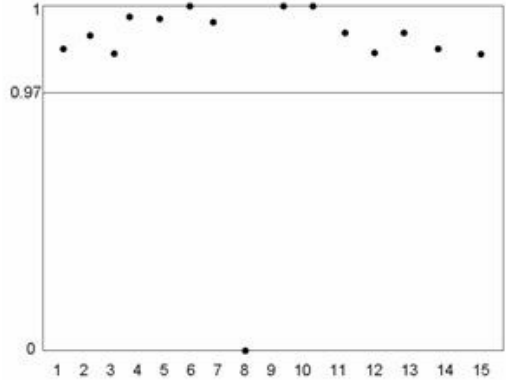

Fig. 4: It depicts the result for proportion of passing of Blum-Blum-Shub PRNG.

**Uniform or Non-uniform distribution of P-values:**

One can have an understanding about uniform or non-uniform distribution of P-values from figs. 5(a) and 5(b). Fig. 5(a) shows a histogram of distribution of P-values for Approximate Entropy Test for BBSG, while Fig. 5(b) shows that for Cumulative Sums Test for BBSG. In both the histograms there are 10 columns: first column indicates the number of P-values lying between 0 and 0.1; second column indicates the number of P-values lying between 0.1 and 0.2, so on and so forth. It may be noted that a methodology is mentioned in the NIST document to calculate the P-value of P-values (POP), where it is stated that P-values for a particular test can be considered uniformly distributed, if it's POP $\geq 0.0001$. For fig. 5(a) and fig. 5(b), the respective POPs are 9.157321e-01 and 8.425709e-07. It is to be noted that P-values are seen to be uniformly distributed in fig. 5(a) for POP = 0.9157, while for POP = 0.000000843 fig. 5(b) shows non-uniform distribution of P-values, although both the tests passed the proportion of passing criterion. From various observations on test results, it is also understood that the larger the POP, the more uniform would be the distribution of P-values.

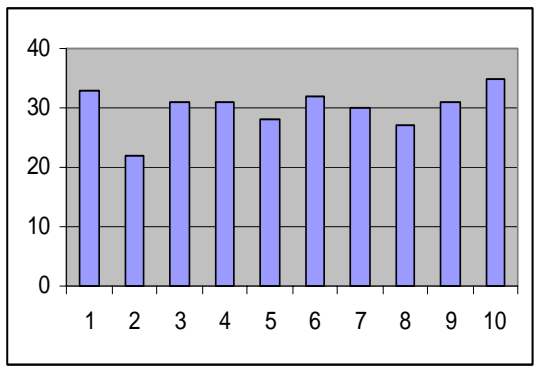
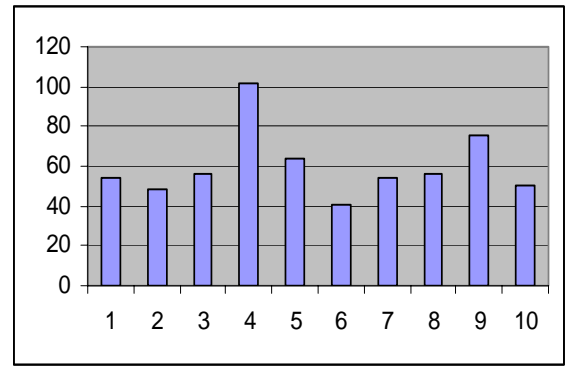

Fig. 5(a). Uniform distribution of P-values of Approximate Entropy test on BBSG.

Fig. 5(b). Non-uniform distribution of P-values of Cumulative Sums test on BBSG.

The results of all tests for all the three algorithms are shown in Appendices (I – III). From this result shown in the Appendices (I – III) one can observe that – LCG by Knuth is unsuccessful for four tests 8, 11, 14 and 15; BBSG is unsuccessful for only test 8; LCG by Park & Miller is unsuccessful for five tests 2, 8, 11, 14 and 15. Considering the proportion of passing as stated above, it is evident that BBSG is a better algorithm than the other two.

However, a study on uniform distribution of POPs for each of the 15 tests is undertaken for all the generators. From the study it is observed that POPs are uniformly distributed for 9 tests of BBSG, while the corresponding figure for Knuth and Park-Miller algorithms is 6 and 7 respectively. Hence, from the consideration of uniformity of POPs, the BBSG is also better than the other two.

**5. Conclusion:**

The package seems to be a perfect one since BBSG is observed to be a better one than the two LCGs. This is also the belief among the cryptographic community. The indigenous code of Test 8 needs a special attention, since all the bit sequences obtained from the three PRNG algorithms do not pass the test, although the bit sequences given in the respective examples of the NIST document do pass the test. It is believed that the test may reject sequences which have too many or too few occurrences of m-runs of ones [1].

**6. Acknowledgement:**

We are thankful to CRSI for providing the support to transform the C codes to a computer package. We are indeed thankful to Prof. Bimal Roy for providing a tacit indication towards indigenous codes. We are grateful to UGC for providing financial support to one of the authors, namely Mr. J K M Sadique Uz Zaman.

**7. A Proposal**

We would like to give the executable package to CRSI so that all who are interested to run NIST Tests can get it from CRSI. In another bundle we would also like to give to CRSI the raw C-codes of the 15 NIST tests developed by us so that interested researchers who intend to look through the respective codes and to study the same can do. The purpose of the proposition is to stimulate intellectual interaction followed by deliberation among Indian researchers working in the field of Cryptology.

## APPENDIX I : Testing of Knuth Algorithm

**Counting of P-values lying in the given ranges**

| Test | 0-.01 | .01-.1 | .1-.2 | .2-.3 | .3-.4 | .4-.5 | .5-.6 | .6-.7 | .7-.8 | .8-.9 | .9-1 |
|---|---|---|---|---|---|---|---|---|---|---|---|
| 1 | 5 | 34 | 29 | 26 | 20 | 30 | 37 | 47 | 0 | 43 | 29 |
| 2 | 6 | 48 | 44 | 30 | 35 | 26 | 30 | 22 | 16 | 26 | 17 |
| 3 | 2 | 31 | 24 | 28 | 26 | 35 | 32 | 41 | 21 | 38 | 22 |
| 4 | 4 | 30 | 33 | 37 | 33 | 17 | 42 | 31 | 21 | 32 | 20 |
| 5 | 3 | 24 | 42 | 24 | 27 | 40 | 20 | 37 | 28 | 29 | 26 |
| 6 | 1 | 6 | 6 | 23 | 27 | 35 | 52 | 50 | 0 | 69 | 31 |
| 7 | 3 | 29 | 36 | 29 | 29 | 20 | 34 | 32 | 27 | 29 | 32 |
| 8 | 300 | 0 | 0 | 0 | 0 | 0 | 0 | 0 | 0 | 0 | 0 |
| 9 | 0 | 0 | 0 | 0 | 0 | 0 | 0 | 0 | 0 | 0 | 300 |
| 10 | 2 | 23 | 28 | 28 | 37 | 40 | 25 | 29 | 27 | 38 | 23 |
| 11 | 313 | 55 | 44 | 38 | 21 | 25 | 24 | 21 | 17 | 23 | 19 |
| 12 | 2 | 32 | 30 | 33 | 37 | 36 | 19 | 20 | 29 | 29 | 33 |
| 13 | 6 | 81 | 63 | 59 | 67 | 44 | 61 | 42 | 40 | 87 | 50 |
| 14 | 156 | 275 | 181 | 187 | 170 | 185 | 196 | 200 | 227 | 245 | 378 |
| 15 | 148 | 296 | 406 | 481 | 520 | 598 | 637 | 640 | 630 | 586 | 458 |

**Status for Proportion of Passing and Uniformity of distribution**

| Test | Expected Proportion | Observed Proportion | Status for Proportion of passing | P-value of P-values | Status for Uniform/Non-uniform distribution |
|---|---|---|---|---|---|
| 1 | 0.972766 | 0.983333 | Success | 2.335246e-08 | Non-uniform |
| 2 | 0.972766 | 0.980000 | Success | 3.409271e-06 | Non-uniform |
| 3 | 0.972766 | 0.993333 | Success | 1.175425e-01 | Uniform |
| 4 | 0.972766 | 0.986667 | Success | 2.197745e-02 | Uniform |
| 5 | 0.972766 | 0.990000 | Success | 7.564718e-02 | Uniform |
| 6 | 0.972766 | 0.996667 | Success | 0.000000e+00 | Non-uniform |
| 7 | 0.972766 | 0.990000 | Success | 7.529784e-01 | Uniform |
| 8 | 0.972766 | 0.000000 | Unsuccess | 0.000000e+00 | Non-uniform |
| 9 | 0.972766 | 1.000000 | Success | 0.000000e+00 | Non-uniform |
| 10 | 0.972766 | 0.993333 | Success | 2.754287e-01 | Uniform |
| 11 | 0.977814 | 0.478333 | Unsuccess | 0.000000e+00 | Non-uniform |
| 12 | 0.972766 | 0.993333 | Success | 2.490301e-01 | Uniform |
| 13 | 0.977814 | 0.990000 | Success | 1.915241e-06 | Non-uniform |
| 14 | 0.983907 | 0.935000 | Unsuccess | 0.000000e+00 | Non-uniform |
| 15 | 0.985938 | 0.972593 | Unsuccess | 0.000000e+00 | Non-uniform |

# APPENDIX II : Testing of Blum-Blum-Shub Algorithm

**Status for Proportion of Passing and Uniformity of distribution**

| Test | 0-.01 | .01-.1 | .1-.2 | .2-.3 | .3-.4 | .4-.5 | .5-.6 | .6-.7 | .7-.8 | .8-.9 | .9-1 |
|---|---|---|---|---|---|---|---|---|---|---|---|
| 1 | 4 | 18 | 25 | 20 | 34 | 37 | 46 | 40 | 0 | 49 | 27 |
| 2 | 3 | 22 | 27 | 38 | 32 | 37 | 25 | 33 | 35 | 23 | 25 |
| 3 | 5 | 16 | 36 | 23 | 28 | 44 | 31 | 33 | 30 | 23 | 31 |
| 4 | 1 | 18 | 27 | 35 | 24 | 28 | 43 | 24 | 43 | 31 | 26 |
| 5 | 1 | 26 | 22 | 32 | 29 | 32 | 37 | 44 | 31 | 19 | 27 |
| 6 | 0 | 1 | 14 | 24 | 26 | 43 | 34 | 48 | 0 | 75 | 35 |
| 7 | 1 | 13 | 24 | 30 | 29 | 31 | 23 | 30 | 27 | 37 | 55 |
| 8 | 300 | 0 | 0 | 0 | 0 | 0 | 0 | 0 | 0 | 0 | 0 |
| 9 | 0 | 0 | 0 | 0 | 0 | 0 | 0 | 0 | 0 | 0 | 300 |
| 10 | 0 | 32 | 29 | 31 | 29 | 28 | 21 | 28 | 35 | 46 | 21 |
| 11 | 5 | 35 | 47 | 51 | 74 | 60 | 72 | 67 | 49 | 60 | 80 |
| 12 | 5 | 28 | 22 | 31 | 31 | 28 | 32 | 30 | 27 | 31 | 35 |
| 13 | 5 | 49 | 48 | 56 | 102 | 64 | 41 | 54 | 56 | 75 | 50 |
| 14 | 32 | 246 | 241 | 223 | 255 | 226 | 250 | 244 | 235 | 220 | 228 |
| 15 | 74 | 435 | 519 | 495 | 527 | 544 | 533 | 578 | 547 | 578 | 570 |

**Status for Proportion of Passing and Uniformity of distribution**

| Test | Expected Proportion | Observed Proportion | Status for Proportion of passing | P-value of P-values | Status for Uniform/Non-uniform distribution |
|---|---|---|---|---|---|
| 1 | 0.972766 | 0.986667 | Success | 4.122711e-10 | Non-uniform |
| 2 | 0.972766 | 0.990000 | Success | 3.949802e-01 | Uniform |
| 3 | 0.972766 | 0.983333 | Success | 1.152721e-01 | Uniform |
| 4 | 0.972766 | 0.996667 | Success | 2.100264e-02 | Uniform |
| 5 | 0.972766 | 0.996667 | Success | 8.378459e-02 | Uniform |
| 6 | 0.972766 | 1.000000 | Success | 0.000000e+00 | Non-uniform |
| 7 | 0.972766 | 0.996667 | Success | 8.309982e-05 | Non-uniform |
| 8 | 0.972766 | 0.000000 | Unsuccess | 0.000000e+00 | Non-uniform |
| 9 | 0.972766 | 1.000000 | Success | 0.000000e+00 | Non-uniform |
| 10 | 0.972766 | 1.000000 | Success | 8.378459e-02 | Uniform |
| 11 | 0.977814 | 0.991667 | Success | 2.042839e-03 | Uniform |
| 12 | 0.972766 | 0.983333 | Success | 9.157321e-01 | Uniform |
| 13 | 0.977814 | 0.991667 | Success | 8.425709e-07 | Non-uniform |
| 14 | 0.983907 | 0.986667 | Success | 2.226391e-01 | Uniform |
| 15 | 0.985938 | 0.986296 | Success | 1.263433e-01 | Uniform |

# APPENDIX III : Testing of Park & Miller Algorithm

**Status for Proportion of Passing and Uniformity of distribution**

| Test | 0-.01 | .01-.1 | .1-.2 | .2-.3 | .3-.4 | .4-.5 | .5-.6 | .6-.7 | .7-.8 | .8-.9 | .9-1 |
|---|---|---|---|---|---|---|---|---|---|---|---|
| 1 | 8 | 23 | 42 | 26 | 34 | 31 | 33 | 43 | 0 | 43 | 17 |
| 2 | 11 | 44 | 41 | 52 | 32 | 25 | 28 | 19 | 23 | 12 | 13 |
| 3 | 3 | 21 | 46 | 26 | 27 | 22 | 36 | 33 | 28 | 31 | 27 |
| 4 | 3 | 18 | 42 | 23 | 36 | 20 | 39 | 18 | 38 | 35 | 28 |
| 5 | 2 | 27 | 33 | 31 | 27 | 41 | 30 | 28 | 29 | 24 | 28 |
| 6 | 0 | 4 | 11 | 18 | 23 | 41 | 37 | 61 | 0 | 58 | 47 |
| 7 | 8 | 36 | 31 | 29 | 25 | 24 | 26 | 30 | 29 | 34 | 28 |
| 8 | 300 | 0 | 0 | 0 | 0 | 0 | 0 | 0 | 0 | 0 | 0 |
| 9 | 0 | 0 | 0 | 0 | 0 | 0 | 0 | 0 | 0 | 0 | 300 |
| 10 | 5 | 35 | 24 | 20 | 33 | 27 | 33 | 27 | 37 | 34 | 25 |
| 11 | 301 | 27 | 33 | 38 | 29 | 29 | 33 | 24 | 22 | 30 | 34 |
| 12 | 5 | 38 | 36 | 33 | 32 | 23 | 32 | 16 | 25 | 25 | 35 |
| 13 | 12 | 67 | 76 | 58 | 75 | 45 | 45 | 44 | 51 | 73 | 54 |
| 14 | 138 | 274 | 169 | 185 | 167 | 224 | 200 | 202 | 235 | 260 | 346 |
| 15 | 146 | 268 | 349 | 499 | 596 | 659 | 628 | 613 | 617 | 564 | 461 |

**Status for Proportion of Passing and Uniformity of distribution**

| Test | Expected Proportion | Observed Proportion | Status for Proportion of passing | P-value of P-values | Status for Uniform/Non-uniform distribution |
|---|---|---|---|---|---|
| 1 | 0.972766 | 0.973333 | Success | 2.781309e-08 | Non-uniform |
| 2 | 0.972766 | 0.963333 | Unsuccess | 3.393663e-11 | Non-uniform |
| 3 | 0.972766 | 0.990000 | Success | 1.004101e-01 | Uniform |
| 4 | 0.972766 | 0.990000 | Success | 3.896945e-03 | Uniform |
| 5 | 0.972766 | 0.993333 | Success | 7.194751e-01 | Uniform |
| 6 | 0.972766 | 1.000000 | Success | 0.000000e+00 | Non-uniform |
| 7 | 0.972766 | 0.973333 | Success | 3.610308e-01 | Uniform |
| 8 | 0.972766 | 0.000000 | Unsuccess | 0.000000e+00 | Non-uniform |
| 9 | 0.972766 | 1.000000 | Success | 0.000000e+00 | Non-uniform |
| 10 | 0.972766 | 0.983333 | Success | 2.093654e-01 | Uniform |
| 11 | 0.977814 | 0.498333 | Unsuccess | 0.000000e+00 | Non-uniform |
| 12 | 0.972766 | 0.983333 | Success | 3.437463e-02 | Uniform |
| 13 | 0.977814 | 0.980000 | Success | 3.436588e-04 | Uniform |
| 14 | 0.983907 | 0.942500 | Unsuccess | 0.000000e+00 | Non-uniform |
| 15 | 0.985938 | 0.972963 | Unsuccess | 0.000000e+00 | Non-uniform |